%% file: main.tex
\documentclass[runningheads]{llncs}
\usepackage[]{graphicx}
\usepackage[]{color}
\usepackage{alltt}
\usepackage[T1]{fontenc}
\usepackage[utf8]{inputenc}
\usepackage{array}
\usepackage{verbatim}
\usepackage{amsmath}
\usepackage{cases}
\usepackage{float}
\usepackage{xspace}
\usepackage{xcolor}
\usepackage[section]{placeins}
\usepackage{adjustbox}
\usepackage{url}
\usepackage{multirow}
\usepackage{diagbox}
\usepackage{pgfplots} 
\usepackage{lipsum}
\usepackage{caption}
\usepackage{subcaption}
\usepackage{marvosym}
\usepackage[T1]{fontenc}

\usepackage{float}

\floatstyle{plaintop}
\restylefloat{table}

\usepackage[tableposition=top]{caption}

\usepackage{hyperref}
\hypersetup{hidelinks,
            linkcolor = blue} 

\graphicspath{ {images/} }

\begin{document}

\newcommand{\gio}[1]{\textcolor{orange}{\textbf{[GQ] #1}}}
\newcommand{\mer}[1]{\textcolor{cyan}{\textbf{[GM] #1}}}
\newcommand{\lore}[1]{\textcolor{green}{\textbf{[LI] #1}}}
\newcommand{\mar}[1]{\textcolor{green}{\textbf{[MG] #1}}}

\newcommand{\red}[1]{\textcolor{black}{#1}}
\newcommand{\redCR}[1]{\textcolor{black}{#1}}

\newcommand\approach{\textit{Deep\-Thought}\xspace}
\newcommand\astra{\textit{ASTRAEA}\xspace}

\title{\approach: a Reputation and Voting-based Blockchain Oracle}	

\author{
Marco Di Gennaro
\inst{1}
\and Lorenzo Italiano
\inst{1}
\and Giovanni Meroni \inst{1, 2}
\and \\
Giovanni Quattrocchi \inst{1}
}

\authorrunning{M. Di Gennaro, L. Italiano, G. Meroni, G. Quattrocchi}
\institute{Dipartimento di Elettronica, Informazione e Bioingegneria,
\\ Politecnico di Milano, Italy 
\and 
DTU Compute, Technical University of Denmark, Denmark
\email{\{giovanni.quattrocchi,lorenzo.italiano\}@polimi.it}\\
\email{marco1.digennaro@mail.polimi.it}\\
\email{giom@dtu.dk}\\
}

\maketitle              

\begin{abstract}\label{chapt:sum}
\input{sections/abstract}
\end{abstract}
\section{Introduction}\label{sec:intro}
\input{sections/intro}

\section{Blockchain and Oracles}\label{sec:background}

\input{sections/background}

\section{\approach}\label{sec:approach}
\input{sections/approach}

\section{Evaluation}\label{sec:results}
\input{sections/results}

\section{Related Work}\label{sec:related}
\input{sections/related}

\section{Conclusions}\label{chapt:conclusions}
\input{sections/conclusions}

\bibliographystyle{plain}
\bibliography{referencelist}

\end{document}

%% file: sections/abstract.tex

\red{
Thanks to built-in immutability and persistence, the block\-chain is often seen as a promising technology to certify information. However, when the information does not originate from the blockchain itself, its correctness cannot be taken for granted. To address this limitation, blockchain oracles ---services that validate external information before storing it in a blockchain--- were introduced. In particular, when the validation cannot be automated, oracles rely on humans that collaboratively cross-check external information. In this paper, we present \approach, a distributed human-based oracle that combines voting and reputation schemes. An empirical evaluation compares \approach with a state-of-the-art solution and shows that our approach achieves greater resistance to voters corruptions in different configurations.
}

%% file: sections/intro.tex

The Web 2.0 revolution made extremely easy for anyone to publish, search and retrieve information \cite{WEB2}. 
As a consequence, organizations and individuals no longer rely on specific sources of information, such as a news agency. Instead, they typically rely on search engines and social media to collect information which, in turn, they may publish after processing, or simply republish as-is.

However, this change of paradigm has also made extremely easy for inaccurate, incorrect, and sometimes forged information to be spread. 
A clear example of this issue is represented by the so-called "fake news" and their detrimental effect they have on society. 
Thus, being able to easily track the provenance of the information available on the Web and to certify its authenticity becomes paramount. Also, to avoid potential conflicts of interests, and to minimize the risk of corruption, the certification process should be carried out independently by multiple subjects.

To this aim, blockchain-based services are seen as a good candidate. 
Originally intended for exchanging cryptocurrency across untrusted entities, blockchain technology evolved to also support the trusted execution of arbitrary code and the exchange of any data. \redCR{In particular, the blockchain provides immutability and persistence and, as reported by Gartner \cite{gartner}, it can be used to track the provenance of information without a centralized authority that validates it. For example, projects such as ANSAcheck \cite{lacity2022blockchain} and SocialTruth\footnote{See \url{http://www.socialtruth.eu}.} exploit this mechanism to certify if a news article comes from an accredited news agency.}
However, the blockchain alone cannot guarantee the authenticity of information that is not natively created on-chain. Instead, it must be coupled with an oracle.

Oracles are services that link a blockchain with the outside world. In particular, they are responsible for providing off-chain data that can be \emph{trusted}, that is, they come from a reliable source \cite{ORACLE-PROBLEM}. A sub-class of such oracles is represented by distributed and human-based ones, where humans manually check the authenticity of off-chain information. These oracles are suited for checking information that cannot be automatically verified, such as a news article. One of the most famous oracles in this class is represented by ASTRAEA \cite{ASTRAEA}, which makes use of voting to determine the outcome. Other approaches, such as Witnet \cite{WITNET}, rely on the reputation of the users. However, we are not aware of any approach that combines both a voting scheme and the reputation.

In this paper, we present \approach, a protocol derived from ASTRAEA that introduces the notion of reputation to increase the level of trust of the information being provided. 
\red{Compared to \astra, \approach makes use of a different scheme to compute the voting outcome according to the reputation of the voters. Also, \approach implements a different mechanism to reward honest voters and to punish dishonest ones.}
\approach can \red{be exploited by both traditional and blockchain-based services to validate information. For example, a news website can exploit \approach to publish only verified news.}  Based on an empirical evaluation, \approach is more robust than \astra,
and makes the corruption of the voting game very difficult for dishonest users.

The rest of this paper is organized as follows: 
Section 2 provides some key background information, while Section 3 illustrates how DeepThought works.
Section 4 presents the evaluation of \approach, Section 5 surveys the related work and Section 6 concludes the paper.

%% file: sections/background.tex
A blockchain is a distributed, immutable ledger, where transactions are secured and verified using a completely decentralized peer-to-peer network~\cite{BLOCKCHAIN}.  Originally, blockchains were conceived to process monetary transactions without relying on a central trusted entity. Transactions are grouped inside blocks that are created using a decentralized consensus algorithm (e.g., proof of work~\cite{BITCOIN}). Each block contains a digest (i.e., a hash) of the previous block, creating a cryptographic chain of blocks that is very hard to be changed by malicious users. 

\red{Nowadays, multiple general-purpose blockchains (such as Ethereum~\cite{ETH}) exist. Such blockchains allow users to write so-called smart contracts~\cite{SZABO}, computer programs whose code and state are stored in the blockchain~\cite{SMART-CONTRACT}. In this way, developers can write ``decentralized'' applications that are transparent and secure, such as ones dedicated to finance~\cite{chen2020Blockchain}, supply chain~\cite{helo2019Blockchains}, and collectibles~\cite{ali2021introduction}.}

Smart contracts are executed in an isolated and deterministic environment and they cannot access information generated outside of the blockchain~\cite{ORACLE-PROBLEM}. This limitation is needed because, while data generated within the blockchain are easily verifiable (i.e., they are the output of traced user transactions), off-chain data are not and cannot be validated in an automated, general-purpose way. 
 
\red{To address this limitation, \textit{oracles} were introduced. An oracle is essentially a trusted service that connects smart contracts to the external world with validated information \cite{ORACLE-FRAMEWORK}. This allows the implementation of richer blockchain applications that can access, for example, stock market fluctuations or news headlines. 
Oracles can be implemented in different ways, ranging from ones that use only off-chain components, to others that are built with smart contracts themselves. In the first case, off-chain components obtain information in ``traditional'' ways, such as by calling a web service, and post the data to the interested smart contracts. In the second case, the information is validated by a smart contract (which is part of the oracle) before being submitted to the contracts that requested it.}

Oracles can be categorized using four criteria \cite{BLOCKCHAIN-ORACLE-EXPLAINED}: software vs hardware, inbound vs outbound, human-based vs unmanned, and centralized vs decentralized.
\textit{Software oracles} can query online sources of information such as websites, web service APIs, and public databases to supply up-to-date information to smart contracts. \textit{Hardware oracles}, instead, have the goal to send information measured from the physical world. For example, in supply chain management, information about a container (e.g., its temperature and location) can be collected by sensors and notified to a smart contract in charge of tracking goods.
\textit{Inbound oracles} supply smart contracts with external data, whereas \textit{outbound oracles} allow smart contracts to interact with the outside world (e.g., opening a smart lock).
\textit{Human oracles} rely on people with deep knowledge of the domain of interest to manually verify the source of information (e.g., a news article) and feed the smart contract with data. Conversely, \textit{Unmanned oracles} process and verify the information using rules and algorithms tailored for the given domain.

An oracle could be \textit{centralized} or \textit{decentralized}, depending on the number of nodes that validate the information.
Given that one of the major advantages of the blockchain is to remove centralized parties, decentralized oracles are usually preferred to centralized ones~\cite{ORACLE-PROBLEM}. However, since multiple parties have to agree on the outcome, decentralized oracles introduce a consensus problem.
In the literature, two main approaches have been proposed to address this issue.  \textit{Reputation-based systems} rely on information provided by parties with a different reputation that measure their reliability. The reputation is usually increased if the information provided is in line with the majority of the other parties, and decreased otherwise. In \textit{voting-based systems}, the parties vote on the correctness of a piece of information (e.g., if the events discussed in a news article are real) by betting an amount of money. To incentivize honest behavior, if the vote is in line with the majority, the user wins a reward. Otherwise, the initial bet is lost.

%% file: sections/approach.tex
\approach is a decentralized blockchain oracle that allows users to validate plain text statements (e.g., to discriminate if a piece of news is legit or fake) that could be then used by other smart contracts in the blockchain. \approach is the first oracle that combines a voting system derived from ASTRAEA with users' reputations to reward the most honest users and to reduce the risk of corruption caused by adversarial users or lazy voters~\cite{DILEMMA}. \approach is an inbound, software oracle, and it is implemented\footnote{Source code available at \url{https://github.com/deib-polimi/deepthought}} as an Ethereum smart contract written in Solidity. Finally, it is human-based because it is assumed that only humans are able to vote rationally on the validity of a plain text statement. 

\subsection{Users and phases}
\begin{figure}[t]
\centering
\includegraphics[width=\linewidth]{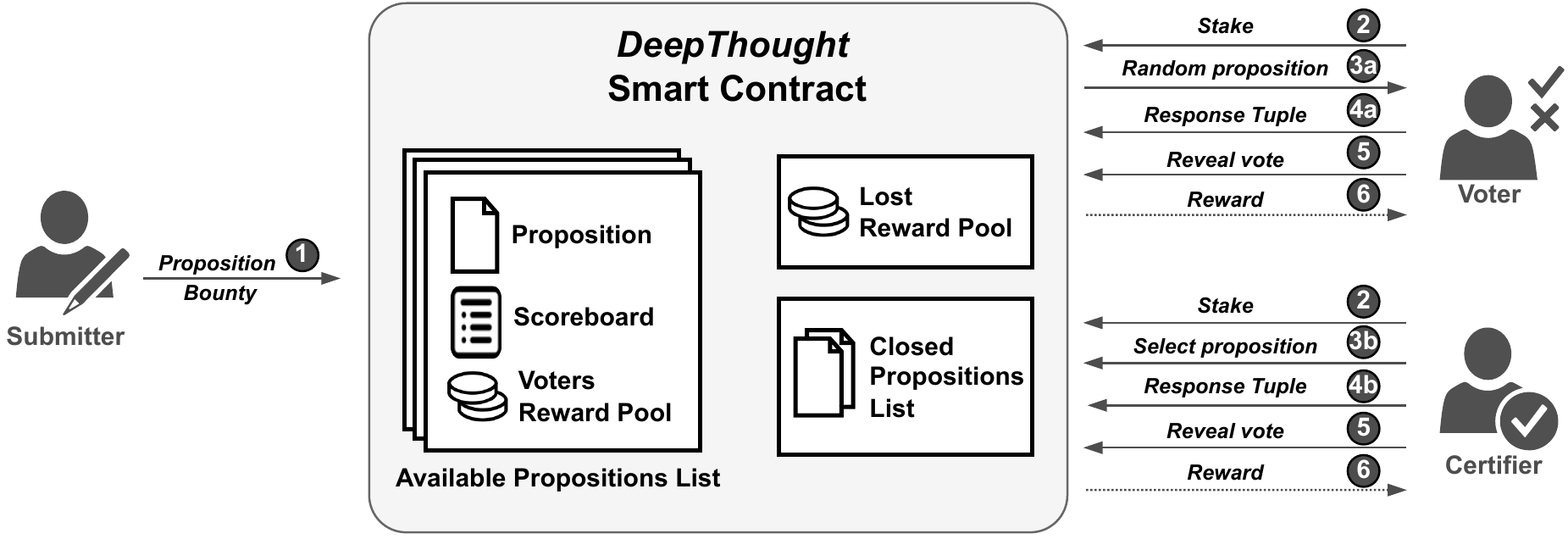}
\caption{\approach overview}
\label{fig:entities}
\end{figure}

Figure \ref{fig:entities} shows a high-level overview of \approach along with all the main entities and interactions. In \approach there are three types of users. 
\vspace{1mm}

\textit{Submitters} are users that publish in the system so-called \textit{propositions}. A proposition is a plain text statement that must be verified through a voting-based process by the other types of users. For example, a proposition could be a piece of news that could be either legitimate or fake. To add a proposition to the system, submitters must pay a so-called \textit{bounty}, a fixed fee that will be used to reward the most honest voters that participated in the verification process.
\vspace{1mm}

\textit{Voters} are users that vote on randomly selected propositions. Voters must vote \texttt{TRUE} if they think that the proposition contains a valid statement, or \texttt{FALSE} otherwise. 
Before submitting a vote on proposition, each voter must \textit{stake} (i.e., lock in the smart contract) an arbitrary amount of money as guarantee for their honesty. After the voting process, if the vote of a voter matches the final outcome and the voter results among the most honest users (more details in the following), the staked amount is sent back to the user along with a reward, otherwise, it is distributed to other users.
\vspace{1mm}

\textit{Certifiers} are users in charge of certifying the outcome of the ballots. To do so, as voters, they vote (either \texttt{TRUE} or \texttt{FALSE}) on the proposition, but unlike voters, they can choose which proposition to work on. For each vote, they have to stake an amount of money which will be returned (plus a reward) to the certifiers only if the final outcome matches their vote. 

\vspace{2mm}

For each proposition, the voting mechanism is organized in six main phases: \textit{submission}, \textit{staking}, \textit{voting}, \textit{certification}, \textit{reveal}, and \textit{closing}. Beforehand, users must subscribe to the system as either submitters, voters, or certifiers. When a new user subscribes, its identifier (i.e., its blockchain address) is stored in the \approach smart contract and its reputation is set to 1.

\vspace{2mm}

\noindent\textbf{Submission phase.} In this phase, a submitter submits a new proposition along with a bounty \red{(action $1$ in Figure~\ref{fig:entities})}. The proposition
is added to the list of \textit{available propositions}, a data structure persisted by the \approach smart contract, and becomes available for the \textit{voting phase}.

\vspace{2mm}

\noindent\textbf{Staking phase.} Before being able to a vote a proposition, each voter and certifier must stake an arbitrary amount of money \red{(action $2$ in Figure~\ref{fig:entities})}. That amount will be regained along with a reward in case of honest and correct votes. Voters must stake an amount $s$ in the range $(min\_s_v, max\_s_v)$ defined 
when the oracle is deployed.
Instead, certifiers stake an amount $s'$ in the range $(min\_c_v, max\_c_v)$  where $min\_c_v > max\_s_v$. In this way, certifiers always stake more money than voters and they are incentivized to pick propositions that they are experts on. The reward sent (eventually) to certifiers is higher than the ones of voters (i.e., the higher the stake, the higher the risk, the higher the reward). In the next phases, each vote is weighted proportionally to the stake and the reputation of the user.

\vspace{2mm}

\noindent\textbf{Voting phase.} 
Once the proposition is submitted, a set of $N$ voters is randomly selected to participate in the voting phase \red{(action $3a$ in Figure~\ref{fig:entities})}. Note that a voter can be randomly selected multiple times for the same proposition and, in this case, will be able to vote more than once.
Being smart contracts deterministic by design, pseudo-random functions are hard to be implemented with them.
In our prototype, we implemented a pseudo-random function that relies on the digest of values that are very difficult, but not impossible, to be predicted. Such values are the timestamp reported in the current blockchain block, and the identifier of the node in the blockchain network that generated the current block.

When the proposition receives $K$ votes, with $K \le N$, the voting phase is concluded. To cast their vote, voters have to indicate the following two values. 
\begin{itemize}
\item The actual \textit{vote}, that is whether the proposition is \texttt{TRUE} or \texttt{FALSE}. 
\item A \textit{prediction value}, which expresses the percentage of how much the voter believes that the outcome will be \texttt{TRUE}. For example, if the voter is certain that a news article is fake because he found some kind of evidence, he will vote \texttt{False} and he will set the prediction value to 0\%.
\end{itemize}

Since transactions in most public blockchains are publicly available, if the votes are stored on-chain in cleartext, malicious and lazy voters could see others' votes and use them for their own interests. For example, they can decide to vote as the majority to obtain a reward, a well-known issue called \textit{herd behavior problem} \cite{HERD-BEHAVIOR}. To avoid this problem, instead of storing the vote on-chain, a digest (i.e., the result of a hashing function) is stored. 

The digest of the vote is computed using the \texttt{keccak256} hashing function.
Also, since a vote can only be \texttt{TRUE} or \texttt{FALSE}, it would be fairly easy for a malicious user to compute all possible outcomes of the hashing function, compare them with the digest of the vote, and then infer the input values. 
To avoid this issue, the hashing function takes as input also a \textit{salt}, which is an arbitrary value selected by the voter and kept secret. 
The digest of the vote 
creates a so-called \textit{response tuple} ($RT$), that submitted to the smart contract \red{(action $4a$ in Figure~\ref{fig:entities})} and  computed 
as follows:
\begin{equation}
\label{eq:voterresponsetuple}
\textit{RT} \, \, = \, \, <kekkak256(\texttt{TRUE}|\texttt{FALSE}, \textit{prediction value}, \textit{salt})>
\end{equation}

\vspace{2mm}

\noindent\textbf{Certification phase.} 
After the proposition has been submitted, and in parallel with the voting phase, certifiers can select the proposition \red{(action $3b$ in Figure~\ref{fig:entities})} submit their votes \red{(action $4b$ in Figure~\ref{fig:entities})} without a prediction value. Their response tuples are computed and stored in the \approach smart contract as follows:

\begin{equation}
\label{eq:certifierresponsetuple}
\textit{RT} \, \, = \, \, <kekkak256(\texttt{TRUE}|\texttt{FALSE}, \textit{salt})>
\end{equation}

The protocol allocates a fixed amount of time (measured in the number of generated blockchain blocks within the smart contract) for the certification phase, since certifiers can decide whether to vote or not vote for a given proposition. Consequently, not all the propositions are guaranteed to have a certification.

\vspace{2mm}

\noindent \textbf{Reveal Phase}. When the certification phase ends, voters and certifiers must reveal their vote by submitting it to the smart contract \red{(action $5$ in Figure~\ref{fig:entities})} along with the \textit{salt} used for computing the digest, and, only for the voters, the prediction value. To ensure that the revealed vote is consistent with the one cast during the voting phase, its digest is computed and compared to the one stored in the smart contract during the voting and certification phases. 

\vspace{2mm}
    
\noindent \textbf{Closing Phase}.
When all voters and certifiers reveal their vote, the proposition is moved from the \textit{list of available propositions} to the \textit{list of closed propositions} and its outcome is computed (either \texttt{TRUE} or \texttt{FALSE} or \texttt{Unknown} in case of a tie). Moreover, as detailed in the following, \approach calculates a score for each voter and creates a \textit{scoreboard} that is used to reward \red{(action $6$ in Figure~\ref{fig:entities})} most honest users. Then, according to the outcome of the proposition, the reputation of each of all the users is updated. Being the votes revealed and stored in the smart contract, any user can check the outcome of the proposition and the voting process.

\subsection{Proposition outcome}

In \approach, the reputation of voters and certifiers is key to compute the outcome of the voting and certification phases.
Intuitively, reputation is used to privilege more reliable users that consistently vote correctly on past propositions.
For each voter and certifier, the user reputation $r$ ranges between $1$ and $max_r$, which can be set during the first deployment of the \approach smart contract. 

For each vote submitted by either a voter and a certifier, a \textit{vote weight} $f$ is computed as follows:

\begin{equation}
\label{eq:voteweight}
f(s, r) = [\alpha\sqrt{s}+(1-\alpha)s]\sqrt{r} 
\end{equation}

where $s$ is the stake submitted along with the vote and $\alpha$ is a value in the range $[0,1]$ defined when the oracle is deployed.
The vote weight is a sub-linear function of the submitted stake. In this way, a single voter is prevented from having dominant voting power~\cite{BRAINS} by using a larger stake. A value of $\alpha$ close to $1$ is used to make the vote weight less dependent on the stake. Conversely, a value of $\alpha$ close to 0 causes the weight to be almost linear with respect to $s$.
Instead, the reputation $r$ scales sub-linearly. This does not affect the outcome between two users with similar reputations, but a user with a high reputation (e.g., 100), is much more significant (e.g., 10) with respect to a newly subscribed one.

Being $\bar{f}_{voters,\texttt{TRUE}}$ the sum of the vote weight $f$ of all the voters that voted \texttt{TRUE} for a given proposition, and $\bar{f}_{voters,\texttt{FALSE}}$ the sum of the vote weight $f$ of all the voters that voted \texttt{FALSE}, the voters outcome $w_{voters}$ is computed as follows: 

 \[ w_{voters}  = 
    \begin{cases} 
       \texttt{TRUE}  &, \ \bar{f}_{voters,\texttt{TRUE}} > \bar{f}_{voters,\texttt{FALSE}} \\
        \texttt{FALSE}  &, \ \bar{f}_{voters,\texttt{TRUE}} < \bar{f}_{voters,\texttt{FALSE}} \\
        \texttt{Unknown}  &, \ \bar{f}_{voters,\texttt{TRUE}} = \bar{f}_{voters,\texttt{FALSE}} \\
    \end{cases}
   \]

The certifiers outcome $w_{certifiers}$ is computed in the same way, by comparing the sum of weights of \texttt{TRUE} and \texttt{FALSE} votes.

\begin{table}
\footnotesize
\caption{Proposition outcome.}
\centering
\label{outcome}
\begin{tabular}{|c|c|c|c|}
\hline
\multicolumn{4}{|c|}{\textbf{Outcome}} \\
\hline
\diagbox{$w_{voters}$}{$w_{certifiers}$} & \texttt{TRUE} & \texttt{FALSE} & \texttt{Unknown} \\
\hline
\texttt{TRUE} & \texttt{TRUE} & \texttt{Unknown} & \texttt{TRUE}  \\
\hline
\texttt{FALSE} & \texttt{Unknown} & \texttt{FALSE} & \texttt{FALSE} \\
\hline
\texttt{Unknown} & \texttt{Unknown} & \texttt{Unknown} & \texttt{Unknown} \\
\hline
\end{tabular}
\end{table}

As shown in Table \ref{outcome}, the outcome of the proposition is computed from $w_{voters}$ and $w_{certifiers}$. If $w_{voters}$ coincides with $w_{certifiers}$, the proposition outcome is the same, otherwise, the outcome is \texttt{Unknown}. In case of the absence of certifiers, or their disagreement, the oracle outcome will be exclusively decided by voters.

Every time a proposition is closed, if the vote of a user matches the outcome the reputation is increased by one, or decreased by one otherwise. Note that a voter may vote multiple times on the same proposition. In this case, the voter is assumed to have voted \texttt{TRUE} (overall) if his/her/their sum of \texttt{TRUE} votes weight is greater than \texttt{FALSE} ones, or vice-versa. The reputation is then modified accordingly.

In the case of an \texttt{Unknown} outcome, the reputation of the voters is not modified, while the reputation of the certifiers is decremented by one.

\subsection{Reward}

After the computation of the proposition outcome, rewards are submitted to a subset of the voters and the certifiers. All the certifiers whose vote matches the outcome of the proposition are rewarded. Voters, instead, are first evaluated using a scoring mechanism, similar to the one reported in \cite{BRAINS}, and only the ones with a high score are rewarded.  

The total score of a voter is based on two different parts:

\begin{itemize}
    \item \textbf{Prediction Score:} A score based on the prediction of the result the voter submitted during the voting phase. $RT_i$ is the response tuple provided by the voter $i$, while $RT_{i'}$ is the response tuple submitted by a randomly selected voter $i'$. $PR$ is the prediction value (i.e., the percentage of how much the voter believes that the outcome will be \texttt{TRUE}) and $IR$ is the vote (either \texttt{TRUE} or \texttt{FALSE}).
    \[ u_{i,PR} = R_q (RT_i.PR, RT_{i'}.IR) \]
    $R_q$ is a quadratic function, given $q$ the prediction and $w$ the outcome, the result will given by:
    \[ R_q(q,w) = 
    \begin{cases} 
        2q - q^2 & , \ w = \texttt{TRUE} \\
        1 - q^2  & , \ w = \texttt{FALSE}
    \end{cases}
                        \]
    \item \textbf{Information Score:} Score based on the information given by the voter.
    \[ u_{i,IR} = 
    \begin{cases}
        1 - (P_{-i,1} - RT_i.PR)^2 & , \ RT_i.IR = \texttt{TRUE} \\
        1 - (P_{-i,0} - RT_i.PR)^2 & , \ RT_i.IR = \texttt{FALSE}
    \end{cases}
    \]
    $P_{-i,q}$ is the arithmetic mean ($G$) of all the $RT.PR$ with $q=RT.IR$ excluding the voter $i$:
    \[ P_{-i,q} = A(RT_q - \{RT_i\}) \]
    \approach uses the arithmetic mean, being the only feasible aggregation mechanism that could be implemented in a smart contract in/with reasonable time/cost. 
\end{itemize}

\noindent The total score of each voter is computed as  

\[ u_i =  u_{i,PR} + u_{i,IR} \]

Each vote is inserted into a scoreboard that is stored in the smart contract. Each vote is added to the scoreboard as soon as it is computed, using algorithm \textit{in-order insertion} in linear time. Given $K$ voters, only the first $K*x$ voters in the scoreboard will earn a reward with $x$ in the range $(0,1)$. In our prototype, we set $x = 0.5$, so that half of the voters are rewarded.

The reward of the voters is computed using function $g_v$ that, as $f$, depends on the stake $s$ and the reputation $r$, and it is computed as follows:
\begin{equation}
    \label{eq:voterreward}
    g_v(s, r) = [\beta s^2+(1-\beta)s]\sqrt{r}
\end{equation}

$\beta$ is a value in the range $[0,1]$ and is used to control the impact of the stake in the reward. The reward is a super-linear function with respect to $s$. On the one hand, this may incentivize high stakes to obtain very high rewards. On the other hand, the scoreboard acts as an opposite force, since only a subset of the voters is rewarded. This also mitigates \textit{Sybil attacks}, where an attacker tries to control the outcome of the oracle by using multiple users~\cite{BRAINS}.  
The reward for voters is taken from a so-called \textit{voters reward pool} that includes the submitter's bounty and all the stakes collected from the voters. The reward is sent to voters starting from the top of the scoreboard. If the funds in the voters reward pool are not enough to reward all the $x*K$ voters, part of them (the lower ranked on the scoreboard) are not rewarded. In the opposite case, when the voters reward pool exceeds the rewards to be distributed, the remaining part is stored in another pool called \textit{lost reward pool}. This pool is shared among all the propositions, and it is used to reward certifiers.

Differently from the voters, the reward for certifiers is always guaranteed if the certification is done correctly (i.e., the vote matches the outcome). Otherwise, their stake is lost and stored in the \textit{lost reward pool}. All the winning certifiers take back their stake and earn a portion (equal for all the certifiers) of the lost reward pool.  

Being $P$ the length of available propositions, $R$ the total amount of money stored in the lost reward pool, and $\bar{s_c}$ the sum of all the certifier stakes that voted correctly on the given proposition, the reward $g_c$ of a certifier depends on the submitted stake $s$ and it is computed as follows:  

\begin{equation}
    \label{eq:certifierreward}
    g_c(s) = s +\frac{R}{P+1}*\frac{s}{ \bar{s_c} }
\end{equation}

In essence, if the certification is correct, the certifier gets back the stake $s$ and a portion of the lost reward pool that is proportional to the staked amount over the certifiers' total staking. By design, the lost reward pool cannot be empty (apart from the very beginning when no proposition has ever been closed), since in the formula its total amount is divided by $P+1$.

%% file: sections/results.tex
%

%

\raggedbottom

Our evaluation focused on the reliability of a decentralized voting oracle, that is, how difficult it is for malicious users to control the outcome of the voting. 
Therefore, our experiments aimed to show if and how much a voting-based oracle extended with a reputation-based system could provide a lower level of corruptibility. In particular, we compared \approach to \astra, 
since the latter was used as the baseline when implementing \approach.

\subsection{Experiment setup}

Although the authors of \astra opted for an analytical evaluation to validate their approach, we opted for an empirical evaluation and, consequently, we implemented a working prototype of \approach. This decision was taken to make easier for other researchers and practitioners to replicate the experiments, as well as to further extend the approach.


To this end, we implemented \astra as an Ethereum smart contract written in Solidity\footnote{Source code available at \url{https://github.com/deib-polimi/deepthought/blob/main/contracts/ASTRAEA.sol}}. We run a set of experiments with different configurations using both \approach implementation and the one developed for \astra. 
All tests were performed using the Truffle Suite\footnote{\url{https://trufflesuite.com}} to simulate an Ethereum Blockchain with 20 active users that interact with the protocol and to deploy the implemented smart contracts.

To validate both \astra and \approach we run a total of 400 experiments, with 10 different configurations and 20 repetitions for each approach. In each configuration, we considered 20 users and 100 different propositions. Each proposition, for the sake of simplicity, was set to have a correct answer equal to \texttt{TRUE}, but we would have obtained the same results with randomized outcomes. Moreover, we split the users into two parts: honest users and adversarial ones. Honest users do not always vote in the correct way. Thus, in each configuration, we varied both the percentage of adversarial users and the accuracy of honest ones. As an example, if the adversarial users are the 25\% of the total and the honest users have an accuracy of 80\%, this means that 75\% of the users will vote \texttt{TRUE} (correct answer) with a probability of $80\%$, while the remaining 25\% will always vote \texttt{FALSE}. These configurations are the ones used in the original \astra evaluation. Thus, our empirical evaluation not only compares \approach and \astra but also validates the analytical assessment of \astra. In particular, we focused on the configurations that involved 20 voters, being more challenging to preserve the robustness of the system in these cases. For the same reason, we also have added four new configurations, which were not presented in the evaluation of \astra, with a higher adversarial control. 

All our tests were performed with a number of voters equals to 20, replicating all specific sub-cases (accuracy equal to 80\% and 95\%, adversary control equal to 0\%, 5\%, 25\%, 35\% and 45\%. The four added configurations focus on an adversarial control equals to 35\% and 45\% while the others were the ones reported in the ASTRAEA evaluation.

\subsection{Results}
\begin{table}[t]
\scriptsize
\caption{Results.}
\begin{adjustbox}{width=\columnwidth,center}
\begin{tabular}{c|cccc|c|ccccc|}
\cline{2-5}
& \multicolumn{4}{|c|}{\textbf{Configuration}} &\multicolumn{6}{c}{\textbf{}}\\
\cline{2-11}
 & \textbf{V} & \textbf{PR} & \textbf{A} & \textbf{ADV} & \textbf{Approach} & \textbf{C-SPEC} & \textbf{C-ANY} & \textbf{STD} & \textbf{MIN} & \textbf{MAX} \\
\hline
\multirow{2}{*}{\#1} & \multirow{2}{*}{20} & \multirow{2}{*}{100}  & \multirow{2}{*}{80}  & \multirow{2}{*}{0}  & \astra & 0.00  & 0.06 & 0.25 & 0 & 1 \\ 
& & & & & \approach & 0.00 & 0.14 & 0.35 & 0 & 1 \\
\hline
\multirow{2}{*}{\#2} & \multirow{2}{*}{20} & \multirow{2}{*}{100}  & \multirow{2}{*}{80}  & \multirow{2}{*}{5}  & 
\astra & 0.00 & 0.26 & 0.63 & 0 & 3 \\
& & & & & \approach & 0.00 & 0.30 & 0.55 & 0 & 2 \\
\hline
\multirow{2}{*}{\#3} & \multirow{2}{*}{20} & \multirow{2}{*}{100}  & \multirow{2}{*}{80}  & \multirow{2}{*}{25}  & 
\astra & 14.29 & 13.88 & 2.73 & 9 & 24 \\
& & & & & \approach & 0.00 & 2.13 & 1.25 & 0 & 4 \\
\hline
\multirow{2}{*}{\#4} & \multirow{2}{*}{20} & \multirow{2}{*}{100}  & \multirow{2}{*}{80}  & \multirow{2}{*}{35}  & 
\astra & 35.00 & 36.40 & 4.12 & 31 & 46 \\
& & & & & \approach & 8.00 & 5.28 & 3.00 & 1 & 13 \\
\hline
\multirow{2}{*}{\#5} & \multirow{2}{*}{20} & \multirow{2}{*}{100}  & \multirow{2}{*}{80}  & \multirow{2}{*}{45}  & 
\astra & 70.00 & 60.60 & 2.39 & 57 & 66 \\
& & & & & \approach & 75.00 & 68.55 & 37.42 & 8 & 99 \\
\hline
\multirow{2}{*}{\#6} & \multirow{2}{*}{20} & \multirow{2}{*}{100}  & \multirow{2}{*}{95}  & \multirow{2}{*}{0}  & 
\astra & 0.00 & 0.00 & 0.00 & 0 & 0 \\
& & & & & \approach & 0.00 & 0.00 & 0.00 & 0 & 0 \\
\hline
\multirow{2}{*}{\#7} & \multirow{2}{*}{20} & \multirow{2}{*}{100}  & \multirow{2}{*}{95}  & \multirow{2}{*}{5}  & 
\astra & 0.00 & 0.00 & 0.00 & 0 & 0 \\
& & & & & \approach & 0.00 & 0.00 & 0.00 & 0 & 0 \\
\hline
\multirow{2}{*}{\#8} & \multirow{2}{*}{20} & \multirow{2}{*}{100}  & \multirow{2}{*}{95}  & \multirow{2}{*}{25}  & 
\astra & 8.00 & 2.40 & 1.35 & 0 & 5 \\
& & & & & \approach & 0.00 & 0.00 & 0.00 & 0 & 0 \\
\hline
\multirow{2}{*}{\#9} & \multirow{2}{*}{20} & \multirow{2}{*}{100}  & \multirow{2}{*}{95}  & \multirow{2}{*}{35}  & 
\astra & 5.00 & 12.90 & 2.31 & 9 & 19 \\
& & & & & \approach & 0.00 & 0.15 & 0.36 & 0 & 1 \\
\hline
\multirow{2}{*}{\#10} & \multirow{2}{*}{20} & \multirow{2}{*}{100}  & \multirow{2}{*}{95}  & \multirow{2}{*}{45}  & 
\astra & 35.00 & 37.70 & 2.74 & 32 & 45 \\
& & & & & \approach & 5.00 & 1.15 & 1.08 & 0 & 3 \\
\hline
\end{tabular}
\end{adjustbox}
\label{tab:results}
\end{table}

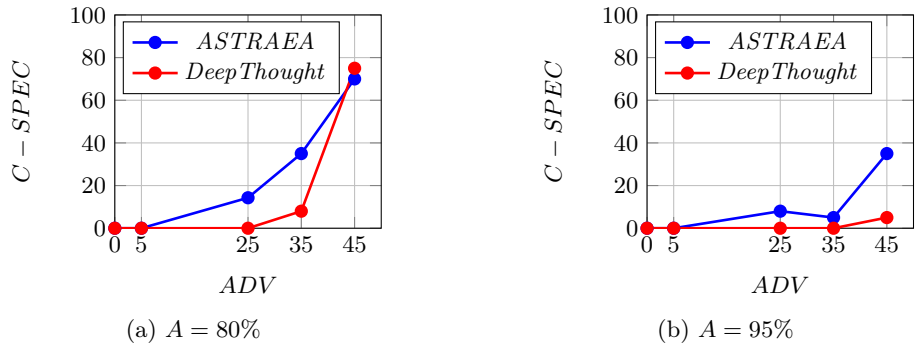
\begin{figure}[t]
    \centering
    \begin{subfigure}[b]{0.42\textwidth}
        \begin{tikzpicture}
        \begin{axis}[
        xmin = 0, xmax = 50,
        ymin = 0, ymax = 100,
        grid = major,
        xlabel = $ADV$, ylabel = $C-SPEC$,
        xtick = {0,5,25,35,45},
        width=\textwidth,
        legend pos=north west
        ]
        \legend{\textsc{\astra},
        \textsc{\approach}}
        
        \addplot [mark=*, color=blue, line width=1pt] coordinates {
        (0, 0)
        (5, 0)
        (25, 14.29)
        (35, 35)
        (45,70)
        };
        \addplot [mark=*, color=red, line width=1pt] coordinates {
        (0, 0)
        (5, 0)
        (25, 0)
        (35, 8)
        (45, 75)
        };
        \end{axis}
        \end{tikzpicture}
         \centering
        \caption{$A = 80\%$}
        \label{fig:cspec80graphic}
    \end{subfigure}
    \hfill
    \begin{subfigure}[b]{0.42\textwidth}
        \centering
        \begin{tikzpicture}
        \begin{axis}[
        xmin = 0, xmax = 50,
        ymin = 0, ymax = 100,
        grid = major,
        xlabel = $ADV$, ylabel = $C-SPEC$,
        xtick = {0,5,25,35,45},
        width=\textwidth,
        legend pos=north west
        ]
        \legend{\textsc{\astra},
        \textsc{\approach}}
        
        \addplot [mark=*, color=blue, line width=1pt] coordinates {
        (0, 0)
        (5, 0)
        (25, 8)
        (35, 5)
        (45, 35)
        };
        \addplot [mark=*, color=red, line width=1pt] coordinates {
        (0, 0)
        (5, 0)
        (25, 0)
        (35, 0)
        (45, 5)
        };
        \end{axis}
        \end{tikzpicture}
         \centering
        \caption{$A = 95\%$}
        \label{fig:cspec95graphic}
    \end{subfigure}
    \caption{Corruptibility of a specific  proposition (C-SPEC).}
    \label{cspecgraphics}
\end{figure}

\begin{figure}[t]
    \centering
    \begin{subfigure}[b]{0.42\textwidth}
        \centering
        \begin{tikzpicture}
        \begin{axis}[
        xmin = 0, xmax = 50,
        ymin = 0, ymax = 100,
        grid = major,
        xlabel = $ADV$, ylabel = $C-ANY$,
        xtick = {0,5,25,35,45},
        width=\textwidth,
        legend pos=north west
        ]
        \legend{\textsc{\astra},
        \textsc{\approach}}
        \addplot [mark=*, color=blue, line width=1pt] coordinates {
        (0, 0.06667)
        (5, 0.2683)
        (25, 13.8857)
        (35, 36.4)
        (45, 60.60)
        };
        \addplot [mark=*, color=red, line width=1pt] coordinates {
        (0, 0.1429)
        (5, 0.3043)
        (25, 2.1304)
        (35, 5.28)
        (45, 68.55)
        };
        \end{axis}
        \end{tikzpicture}
        \centering
        \caption{$A = 80\%$}
        \label{fig:cany80graphic}
    \end{subfigure}
    \hfill
    \begin{subfigure}[b]{0.42\textwidth}
        \centering
        \begin{tikzpicture}
        \begin{axis}[
        xmin = 0, xmax = 50,
        ymin = 0, ymax = 100,
        grid = major,
        xlabel = $ADV$, ylabel = $C-ANY$,
        xtick = {0,5,25,35,45},
        width=\textwidth,
        legend pos=north west
        ]
        \legend{\textsc{\astra},
        \textsc{\approach}}
        
        \addplot [mark=*, color=blue, line width=1pt] coordinates {
        (0, 0)
        (5, 0)
        (25, 2.4)
        (35, 12.9)
        (45, 37.70)
        };
        \addplot [mark=*, color=red, line width=1pt] coordinates {
        (0, 0)
        (5, 0)
        (25, 0)
        (35, 0.15)
        (45, 1.15)
        };
        \end{axis}
        \end{tikzpicture}
        \caption{$A = 95\%$}
        \label{fig:cany95graphic}
    \end{subfigure}
    \caption{Corruptibility of ANY proposition (C-ANY).}
    \label{canygraphics}
\end{figure}
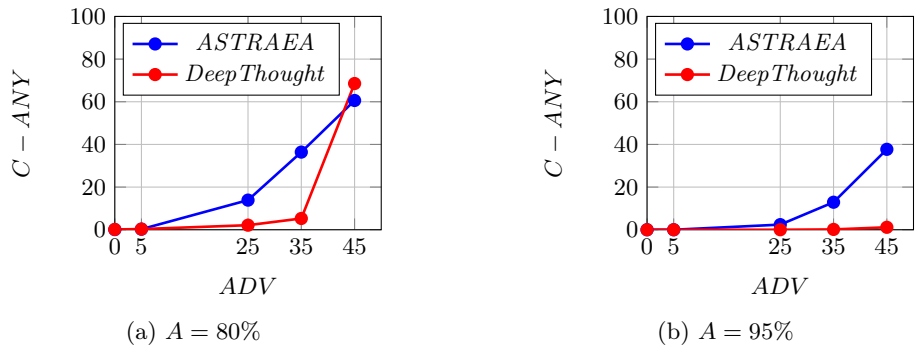

Table \ref{tab:results} shows the results obtained by \approach and \astra in the ten tested configurations.
If we compare the behavior of \astra with the assessment reported in \cite{ASTRAEA} (configurations \#1, \#2, \#3, \#6, \#7 and \#8), the results are comparable except for some smaller deviations that could have been introduced by statistical errors. In this way, we also made sure that our implementation of \astra is correct.

For each configuration, the table shows the number of voters (V), the number of proposition (PR), the accuracy of voters (A, in percentage), the adversarial control (ADV, in percentage), the percentage of times that a proposition selected by an adversarial user was corrupted (C-SPEC), the average number of corrupted propositions (C-ANY) along with the standard deviation (STD), minimum (MIN), and maximum (MAX) values. It must be noted that we considered corrupted also the propositions whose outcomes were incorrect for mistakes made by honest but inaccurate users.

In the configurations with ADV less than or equal to 25\% \approach obtained zero cases where C-SPEC is greater than $0$. On contrary, \astra obtained $14.29\%$ and $8\%$ in configurations \#3 and \#8 respectively where the adversary control is equal to $25\%$. These data seem to indicate that \approach is more robust compared to \astra thanks to the usage of reputation.

Configuration \#1 and \#2 show that \astra obtain a smaller amount of corruptions on average compared to \approach. It must be noted that in these configurations, the adversarial control is quite small, 0\% and 5\% respectively. This shows that \approach is more sensible than \astra to the accuracy of honest voters. Intuitively, if voters with a higher reputation make mistakes, the chances of a wrong output are slightly higher compared to \astra. On this note, it must be noted that in the experiment the accuracy is equal to all the voters. However, in a real-world scenario voters with a higher reputation should tend to make fewer mistakes compared to others and make \approach perform better than \astra also in these cases. This behavior is also confirmed by configurations \#6 and \#7 where the adversarial control is equal to \#1 and \#2 respectively and \approach obtains zero corruptions in all the repetitions. 

Configurations \#3 and \#8 show a significant difference between the behavior of \approach and \astra. These cases are quite challenging, since one-fourth of the voters are adversarial. In these cases, \approach obtained only 2.13 and 0.0 corrupted propositions on average, compared to the much higher data of \astra 13.88 and 2.4 respectively. These results clearly show the benefit of \approach. 

\red{Configurations \#4, and \#9  show a huge difference between \approach and \astra. In these newly introduced configurations, the adversarial control reaches 35\%. In configuration \#4 \approach is able to keep C-SPEC as low as 8\% and obtained 5.28 corrupted propositions on average, while in configuration \#5 (with a higher voters accuracy) the results are even lower with C-SPEC and C-ANY equal to 0\%  and 0.15 respectively. On contrary, \astra shows a very high level of corruptibility with C-SPEC and C-ANY equal to 35\% and 36.40 in configuration \#4 and 5\% and 12.90 in configuration \#5. A similar difference is obtained in configuration \#10 where the adversarial control is 45\% and the accuracy of voters is very high (95\%). This shows how \approach is robust also in very edge cases when almost half of the total of voters is dishonest.}

\red{When the adversarial control is very high (45\%) and the accuracy is lower (80\%), the performance of our approach drops significantly (configuration \#5). In this case, both \astra and \approach are not able to keep the system under control with a very high probability of corruption of a specific proposition (> 70\%) and more than 60 propositions corrupted on average. In this case, \approach shows slightly worse results compared to \astra. Our approach is more sensitive to the initial voting rounds, if honest voters outperform dishonest ones the system remains under control (MIN equals 8) since honest voters accumulate reputation. However, if dishonest voters are able to corrupt the initial propositions, their reputation increases and the low accuracy of honest voters is not enough to keep the system reliable (MAX equals 99). This behavior is also captured by the very high standard deviation of \approach (37.42) compared to \astra's one (2.39).}

Figures~\ref{cspecgraphics} and \ref{canygraphics} help visualize the different behaviors of \approach and \astra when the adversarial control increases. 
By combining a voting-based approach with a reputation-based system, our solution is significantly more robust than \astra in avoiding corruptions of specific propositions (Figure \ref{cspecgraphics}) and on average (Figure \ref{canygraphics}) in almost all the cases.

%% file: sections/related.tex

As already mentioned, one of the most successful voting-based oracles is represented by \astra \cite{ASTRAEA}. Similarly to \approach, \astra also relies on submitters, voters and certifiers. However, it relies on a different scheme to determine the voting outcome and to compute the rewards. In particular, each vote has the same weight, regardless of who cast it. Also, two  pools, containing the bounty of all the voters and certifiers who voted, respectively \texttt{TRUE} and \texttt{FALSE}, are used to compute the reward. However, these choices make \astra not so robust to adversary control. Similarly, \astra can be subject to the \textit{Verifier's Dilemma}, that is,  users always voting and certifying with a constant value in order to maximize their profit without expending any effort \cite{DILEMMA}.

To address these limitations, several extensions of \astra have been proposed in the literature. 
%
Shintaku \cite{SHINTAKU} removes the role of the certifier, leaving the certification of the voting result to the voters themselves. To counter the \textit{Verifier's Dilemma}, a voter has to answer a pair of randomly selected propositions, instead of a single one. Voters are then eligible for rewards only if their votes for the two propositions differ. However, this approach has been criticized in \cite{PAIREDQUEST} and \cite{BRAINS} for being practically ineffective against lazy voting unless the penalties for disagreement are at least twice as large as rewards for agreement. Moreover, the honest voters are not incentivized because the payoffs are low.

Merlini et al. in \cite{PAIREDQUEST} require a submitter to submit two antithetic propositions, posting a bond. Once votes are collected, the oracle checks whether the two questions
converged to different answers. 
%
%
Cai et al. in \cite{BRAINS} introduce a non-linear scoring scheme to weight the votes and compute the rewards. In particular, this approach collects for each vote a binary information answer and a popularity prediction. The oracle answer is determined by the majority of the information answer, weighted by the associated stakes and adjusted by a sub-linear function. Then, the oracle assigns a score to each report based on the accuracy and the degree of agreement with peers. Only the top-scored voters are awarded, while the share of award is determined by their stake adjusted by a super-linear function.

With respect to these approaches, \approach introduces the concept of reputation, which is absent in all of them. The reputation is used to determine the answer, the rewards and the penalties using a scoring scheme similar to \cite{BRAINS}.



Regarding reputation-based oracles, the most famous is Witnet \cite{WITNET}. This oracle runs on its own native customized blockchain, which provides support for smart contracts and relies on tokens named Wit. Miners, that are called \textit{witnesses}, can earn Wits by retrieving and validating external information to be inserted into smart contracts. Witnesses contribute with their mining power, which is mainly determined by their reputation. Similar to \approach, Witnet rewards the successful majority consensus witnesses, while penalizing the contradicting witnesses. However, it does not have certifiers to counter-check the outcome of voters.

%

%% file: sections/conclusions.tex
In this paper, we presented \approach, a decentralized human-based oracle that combines voting with reputation. \redCR{\approach seeks to demonstrate how a reputation-weighted voting system could decrease the probability of outcome corruption compared to existing solutions available in the state-of-the-art that only rely on simpler voting mechanism or only on users' reputation.} The results of the empirical evaluation, \redCR{carried out through the implementation of two smart contracts}, show that \approach presents a higher resistance to adversary control than \astra, which relies only on a voting scheme. In the future, we will implement other voting-based oracles available in the literature and compare them with \approach. \redCR{Moreover, we will extend our evaluation with the assessment of a real-world use case and our reward mechanism. }